\documentclass[showpacs,preprintnumbers,amsmath,amssymb,nofootinbib,floatfix]{revtex4}

\usepackage[dvips]{graphicx}
\usepackage{dcolumn}
\usepackage{bm}

\def\mapgeq{\mathbin{\lower.3ex\hbox{$\buildrel>\over{\smash{\scriptstyle\sim}\vphantom{_x}}$}}}
\def\mapleq{\mathbin{\lower.3ex\hbox{$\buildrel<\over{\smash{\scriptstyle\sim}\vphantom{_x}}$}}}
\def\mapgeqeq{\mathbi{\lower.3ex\hbox{$\buildrel>\over{\smash{\scriptstyle\approx}\vphantom{_2}}$}}}
\def\mapleqeq{\mathbin{\lower.3ex\hbox{$\buildrel<\over{\smash{\scriptstyle\approx}\vphantom{_2}}$}}}
\mathchardef\hanaO="724F
\def\Journal#1#2#3#4{{#1} {\bf #2} (#4) #3}

\def\NPSUPPL{Nucl. Phys. Proc. Suppl.}
\def\PLB{{Phys. Lett.} B}

\def\PRL{Phys. Rev. Lett.}
\def\RMP{Rev. Mod. Phys.}

\def\PTP{Prog. Theor. Phys.}

\def\EPJ{Euro. Phys. J. C}

\def\JETPUSSR{Sov. Phys. JETP}

\def\ZETP{Zh. Eksp. Teor. Fiz.}

\def\SCI{Science}
\def\APJ{Astrophys. J.}

\def\NJP{New J. Phys.}

\def\Erratum{Erratum-ibid}

\begin{document}

 \title{The Double Cover of the Icosahedral Symmetry Group and Quark Mass Textures}

\author{Lisa L.~Everett and Alexander J.~Stuart}
\affiliation{  \vspace{3mm}
Department of Physics, University of Wisconsin, Madison, WI, 53706, USA}

\date{\today}

\begin{abstract}
We investigate the idea that the double cover of the rotational icosahedral symmetry group is the family symmetry group in the quark sector.  The icosahedral ($\mathcal{A}_5$) group was previously proposed as a viable family symmetry group for the leptons.   To incorporate the quarks, it is highly advantageous to extend the group to its double cover,  as in the case of tetrahedral ($A_4$) symmetry.  We provide the basic group theoretical tools for flavor model-building based on the binary icosahedral group $\mathcal{I}'$ and construct a model of the quark masses and mixings that yields many of the successful predictions of the well-known $U(2)$ quark texture models.
\end{abstract}

  \preprint{MADPHYS-10-1566}                            
\maketitle

\section{Introduction}
With the measurement of neutrino oscillations \cite{SK,K2K,OldSolar,Sun,Reactor,Theta13}, an intriguing pattern of lepton mixing has emerged.  The neutrino oscillation data have revealed that two of the mixing angles of the Maki-Nakagawa-Sakata-Pontecorvo (MNSP) \cite{PMNS} mixing matrix are large and the third angle is bounded from above by the Cabibbo angle (for global fits, see \cite{NuData}).   This pattern, with its striking differences from the quark mixing angles of the Cabibbo-Kobayashi-Maskawa (CKM) matrix, has shifted the paradigm for addressing the Standard Model (SM) flavor puzzle.\\ \vspace{-0.14in}

More precisely, while the quark sector previously indicated a flavor model-building framework based on the Froggatt-Nielsen mechanism \cite{FN} with continuous family symmetries (see also \cite{kl}), the large lepton mixing angles suggest discrete non-Abelian family symmetries.  Lepton flavor models  have been constructed based on the tetrahedral symmetry group $\mathcal{T}$, which is isomorphic to $\mathcal{A}_4$, the alternating group of four elements \cite{A4},  the binary tetrahedral group $\mathcal{T}^\prime$ \cite{carone,tprime,ding},  $\Delta(3n^2)$ and $\Delta(6n^2)$ \cite{delta}, the semidirect product of $\mathcal{Z}_3$ and $\mathcal{Z}_7$ \cite{Z3xZ7}, $\mathcal{Z}_2\times \mathcal{Z}_2$ \cite{raidal}, $\mathcal{S}_4$ \cite{S4},  $\mathcal{S}_3$ \cite{S3}, the semidirect product of $\mathcal{A}_4$ and $\mathcal{S}_3$ \cite{S3xA4}, $\mathcal{PSL}(2,7)$ \cite{PSL27},  the quaternionic symmetries \cite{Quat}, the dihedral symmetries $\mathcal{D}_n$ \cite{dihedral,D10},  and the icosahedral symmetry group $I$, which is isomorphic to $\mathcal{A}_5$  \cite{A5} (a four-family $\mathcal{A}_5$ model can be found in \cite{Chen:2010ty}). Recent reviews of discrete group theory and flavor model building can be found in \cite{reviews,rodejohann,ludl}.  
The natural next step is to incorporate the quarks and obtain a complete flavor theory.   Several models include both quarks and leptons, in some cases within a fully grand unified theory.  There are also models that focus only on quark mixing (see e.g.~\cite{discretequarks}).\\ \vspace{-0.14in}

In certain cases, however, the quarks are most easily accommodated if the original group is extended to its double covering group, which allows for spinorial representations.   The most prominent example is the tetrahedral ($\mathcal{A}_4$) group \cite{A4}, which has three one-dimensional representations and one triplet representation.  By considering its double cover, the binary tetrahedral group $\mathcal{T}$,  and assigning the lighter generation quarks to doublets and the third generation quarks to singlets,  the successful $U(2)$ quark textures \cite{u2} can be obtained together with the lepton sector prediction of Harrison-Perkins-Scott [HPS] ``tri-bimaximal" mixing \cite{HPS}, as shown in \cite{carone,tprime,ding}.\\ \vspace{-0.14in}

In this paper, we investigate the extension of the icosahedral symmetry group $\mathcal{I}$ ($\mathcal{A}_5$) to its double cover, the binary icosahedral group $\mathcal{I}^\prime$, and consider $\mathcal{I}^\prime$ as a family symmetry group for the quarks.   This study is based on our previous work \cite{A5} in which we investigated $\mathcal{I}$  as a family symmetry group for the leptons.  The icosahedral symmetry group, which is one of the discrete groups based on the five Platonic solids, had been comparatively unexplored for physics applications at least in part because it is not a crystallographic point group.  In our work, we explored the interesting hypothesis that the solar angle is given by $\tan\theta_{\rm sol}=1/\phi$, where $\phi=(1+\sqrt{5})/2$ is the golden ratio.  This hypothesis was first suggested in \cite{datta} and later explored within a $\mathcal{Z}_2\times\mathcal{Z}_2$ framework in \cite{raidal}, who also provided a prediction for the Cabibbo angle based on the golden ratio that is correct to three digits and suggested $\mathcal{A}_5$ as a useful setting for lepton flavor model building since the golden ratio appears in the geometry of the icosahedron.  Alternative ideas relating the golden ratio and the solar mixing angle include a  $\mathcal{D}_{10}$ model that predicts $\cos\theta_{\rm sol}=\phi/2$ \cite{D10}.\\ \vspace{-0.14in} 

We will see that to incorporate the quarks, it is highly advantageous to extend $\mathcal{I}$ to its double cover $\mathcal{I}^\prime$.  Hence, in this work we provide the group theoretical tools needed for building $U(2)$-inspired quark flavor models based on $\mathcal{I}^\prime$.  As in the $\mathcal{T}^{\prime}$ case, the $U(2)$ textures are expected to be possible within $\mathcal{I}^\prime$, since $\mathcal{I}^\prime$ has doublet representations that obey the basic $U(2)$ relation $\bf{2} \otimes  \bf{2}=\bf{1}  \oplus \bf{3}$.  Therefore, as a working example, we also construct a supersymmetric $\mathcal{I}^\prime \otimes \mathcal{Z}_9$ quark flavor model based on that relation that is similar to the quark sector of the $\mathcal{T}^\prime \otimes \mathcal{Z}_3\otimes \mathcal{Z}_9$ model of \cite{ding}.  In future work, we will provide a more comprehensive overview of $\mathcal{I}^\prime$ and consider flavor model building for both the leptons and quarks together within the $\mathcal{I}^\prime$ framework \cite{alexlisainprep}. \\ \vspace{-0.14in}

The outline for this paper is as follows.  We begin by describing the basic features of the icosahedral symmetry group $\mathcal{I}$ and its double cover $\mathcal{I}^\prime$.   Next, we will outline the group presentation and the construction of group invariants for $\mathcal{I}^\prime$.  We then present our quark flavor model based on $\mathcal{I}^\prime \otimes \mathcal{Z}_9$ and address the issue of vacuum alignment for the flavon sector.  Finally, we present our conclusions and outlook.

\section{Theoretical Background}

We begin with a brief overview of the icosahedral symmetry group $\mathcal{I}$ and its double cover, the binary icosahedral group $\mathcal{I}^\prime$.  The basic theory of the icosahedral symmetry group is known in the mathematical literature, and can be found in the papers \cite{backhousegard,cumminspatera,shirai,hoyle,luhn,A5}.  Finite groups are also covered extensively in several texts \cite{Coxeter,Hamermesh,Lomont,Ramond}, and reviews \cite{ludl}.\\ \vspace{-0.14in}

The icosahedral symmetry group $\mathcal{I}$ is the group of all rotations that preserve the orientation of the icosahedron, which is the Platonic solid consisting of twenty equilateral triangles.  (Here and in what follows we will only consider proper rotations, and therefore ignore inversions.)  $\mathcal{I}$ is the finite subgroup of $SO(3)$ that is isomorphic to $\mathcal{A}_5$, the alternating group of five elements.   $\mathcal{I}$ contains 60 elements:  the identity, rotations by $2\pi/5$ and  $4\pi/5$  about an axis through each of the twelve vertices, rotations by $2\pi/3$ about an axis through the center of each of the twenty faces,  and rotations by $\pi$ about the midpoint of each of the thirty edges (resulting in fifteen distinct rotations).  These five types of rotations form five conjugacy classes, which are denoted as follows:
\begin{eqnarray}
1, \, 12C_5, \, 12C_5^2, \, 20C_3, \, 15C_2.
\end{eqnarray}
Here we follow the standard procedure and use Schoenflies notation: $C_n^k$ is a rotation by $2k\pi/n$, and the number in front gives the number of group elements in the conjugacy class. The elements of $C_n^k$ are known as order $n$ elements, {\it i.e.}, they result in the identity after $n$ operations.   Given the basic results in discrete group theory that the number of elements in the group is equal to the sum of the squares of the irreducible representations, and that the number of conjugacy classes is equal to the number of irreducible representations, the 60 elements of $\mathcal{I}$ result in the condition:
\begin{eqnarray}
 1+12+12+15+20=60=1^2+3^2+3^2+4^2+5^2.
\end{eqnarray}
Thus, $\mathcal{I}$ has five irreducible representations: ${\bf 1}$, ${\bf 3}$, ${\bf 3}^\prime$, ${\bf 4}$, and ${\bf 5}$.  The presence of the two distinct triplet representations suggests that $\mathcal{I}$ may be a good candidate for a family symmetry group, as explored in our previous investigation \cite{A5}.  We note here that in contrast to $\mathcal{T}^\prime$, which has three one-dimensional representations, the only one-dimensional representation of $\mathcal{I}$ is the identity.  This fact, which will be of significance when we turn to the construction of flavor models, results because $\mathcal{I}$ is a perfect group; {\it i.e.}, it is equal to its commutator subgroup \cite{Lomont,Ramond}. \\ \vspace{-0.14in}

The double cover of the icosahedral group, which is denoted as the binary icosahedral group $\mathcal{I}^\prime$, has double the number of group elements as $\mathcal{I}$, for a total of 120.  The group elements are grouped into the five conjugacy classes of $\mathcal{I}$ and four additional conjugacy classes:
\begin{eqnarray}
R, \, 12C_5 R, \, 12C_5^2 R, \, 20C_3 R,
\end{eqnarray}
where $R$ is $-I$, and $I$ denotes the identity.  The absence of another copy of $15 C_2$ occurs because the additional group elements in $\mathcal{I}^\prime$ double the order of the existing $C_2$ conjugacy class from 15 to 30, instead of forming a new conjugacy class.  This information can be used to calculate the additional irreducible representations of ${\mathcal{I}^{\prime}}$:
\begin{eqnarray}
1+12+12+20+15=60=2^2+2^2+4^2+6^2.
\end{eqnarray}
Hence, $\mathcal{I}^\prime$ has the spinorial representations  ${\bf 2}$, ${\bf 2}^\prime$, ${\bf 4}^\prime$, and ${\bf 6}$ in addition to the ${\bf 1}$, ${\bf 3}$, ${\bf 3}^\prime$, ${\bf 4}$, and ${\bf 5}$ representations.\\ \vspace{-0.14in}   

The character table of $\mathcal{I}^\prime$, which can also be found in \cite{cumminspatera,shirai,luhn}, is given in Table~\ref{chartable}.  We see that the golden ratio, $\phi=(1+\sqrt{5})/2$ (where $\phi-1=1/\phi$), appears for the order five elements for the doublet and triplet representations. \\ \vspace{-0.14in}

\begin{table}[tbp]
\begin{tabular}{|c|c|c|c|c|c||c|c|c|c|}
\hline
$\mathcal{I^{\prime}}$& $\;$$\;$ \textbf{1} $\;$$\;$&\textbf{3}&$\textbf{3}^{\prime}$& $\;$$\;$ \textbf{4} $\;$$\;$ & $\;$$\;$ \textbf{5} $\;$$\;$ &\textbf{2}&\textbf{2}$^{\prime}$& $\;$$\;$ \textbf{4}$^{\prime}$ $\;$$\;$&$\;$ $\;$ \textbf{6}$\;$ $\;$ \\ \hline
\textbf{$1$}& 1 &3&3&4&5&2&2&4&6\\ \hline
\textbf{$12 C_5$}&1&$\phi$&$1-\phi$&$-$1&0&$\phi$&$1-\phi$&1&$-$1\\ \hline
$12{C}^2_{5}$&1&$1-\phi$&$\phi$&$-$1&0&$\phi -1$&$-\phi$&$-$1&1\\ \hline
$20{C}_{3}$&1&0&0&1&$-$1&1&1&$-$1&0\\ \hline
$30 {C}_{2}$&1&$-$1&$-$1&0&1&0&0&0&0 \\ \hline
\hline
$R$&1&3&3&4&5&$-$2&$-$2&$-$4&$-$6\\ \hline
$12 C_5R$&1&$\phi$&$1-\phi$&$-$1&0&$-\phi$&$\phi -1$&$-$1&1\\ \hline
$12 C_5^2R$&1&$1-\phi$&$\phi$&$-$1&0&$1-\phi$&$\phi$&1&$-$1\\ \hline
$20C_3R$&1&0&0&1&$-$1&$-$1&$-$1&1&0\\ \hline
\end{tabular}
\caption{\label{chartable}The character table of the binary icosahedral group $\mathcal{I^{\prime}}$, in which  $\phi$ is the golden ratio: $\phi=(1 + \sqrt{5})/2$.}
\end{table}

From the character table, it is straightforward to deduce the form of the tensor products of the irreducible representations, which are known in the literature \cite{cumminspatera,shirai,luhn}.  The tensor products involving the doublet and triplet representations are as follows (for the complete list, see \cite{cumminspatera,shirai,luhn}):
\begin{eqnarray}
{\bf 3}\otimes {\bf 3} &=& {\bf 1}\oplus {\bf 3}\oplus {\bf 5}, \qquad {\bf 3}^\prime \otimes {\bf 3}^\prime = {\bf 1}\oplus {\bf 3}^\prime \oplus {\bf 5}, \qquad \;\;\; {\bf 3}\otimes {\bf 3}^\prime = {\bf 4}\oplus {\bf 5},\nonumber \\
{\bf 2}\otimes {\bf 2} &=& {\bf 1}\oplus {\bf 3}, \qquad \;\;\;\, \;\;\;{\bf 2}^\prime \otimes {\bf 2}^\prime = {\bf 1}\oplus {\bf 3}^\prime, \qquad  \;\;\;\, \;\;\;\;\;\; {\bf 2}\otimes {\bf 2}^\prime = {\bf 4},\nonumber \\
{\bf 2}\otimes {\bf 3} &=& {\bf 2}\oplus {\bf 4}^\prime,
\;\;\;\;\;\, \qquad {\bf 2}^\prime \otimes {\bf 3}^\prime = {\bf 2}^\prime \oplus {\bf 4}^\prime  \qquad  \;\;\;\;\;\;\;\;\;\; {\bf 2}\otimes {\bf 3}^\prime = {\bf 6},\;\;{\bf 2}^\prime \otimes {\bf 3}={\bf 6}.
\end{eqnarray}
Observe that the basic $U(2)$ relation, ${\bf 2}\otimes {\bf 2} = {\bf 1}\oplus {\bf 3}$ (and an analogous relation involving the primed representations), holds within the $\mathcal{I}^\prime$ group.  In what follows, we will discuss the construction of quark flavor models based on this relation in analogy with the well-known $U(2)$ flavor models \cite{u2}.  First, however, we will need the $\mathcal{I}^\prime$ group presentation and the construction of group invariants, which is the subject of the next section.

\section{Group Presentation and Group Invariants}
The elements of discrete groups such as the binary icosahedral group can be generated by a set of basic elements that satisfy certain relations; the elements and the rules together are known as the ``presentation" of the group.  For the case of $\mathcal{I}$ and $\mathcal{I}^\prime$, there are several equivalent group presentations (see e.g.~\cite{cumminspatera,luhn,shirai,hoyle}).  Following our previous work on $\mathcal{I}$ ($\mathcal{A}_5$) \cite{A5}, we prefer to use the presentation of Shirai \cite{shirai}, in which all group elements can be expressed in terms of the order two generator $S$ and the order 5 generator $T$, such that
\begin{eqnarray}
 S,\;T:  S^2=T^5=R, \qquad (T^2ST^3ST^{-1}STST^{-1})^3=I.
 \end{eqnarray}
 Here $R$ is the identity $I$ in the case of the single-valued representations ${\bf 1}$, ${\bf 3}$, ${\bf 3}^\prime$, and ${\bf 5}$, and $R$ is $-I$ for the double-valued representations ${\bf 2}$, ${\bf 2}^\prime$, ${\bf 4}^\prime$, and ${\bf 6}$.  We note that while $S$ and $T$ are contained in the conjugacy classes $C_2$ and $C_5$, respectively, the order three combination of $S$ and $T$ given above is contained in the $C_3R$ conjugacy class (not $C_3$),  which is why it cubes to the identity rather than $R$. \\ \vspace{-0.14in}
 
 The Shirai presentation can be related to other presentations that are standard in the literature, such as the presentation of Threlfall involving order three and order five generators found for example in \cite{Coxeter}:
 \begin{eqnarray}
 a,\; b: a^3=b^5=(ab)^2=R,
 \end{eqnarray} 
 and the presentation described in the work of Cummins and Patera \cite{cumminspatera}, which involves two generators of order two and one order three generator:
 \begin{eqnarray}
 A_1,\; A_2, \; A_3: A_1^3=A_2^2=A_3^2=R, \qquad (A_1 A_2)^3=(A_2A_3)^2=(A_1A_3)^3=R.
 \end{eqnarray} 
The generators of the Shirai presentation can be related  (up to similarity transformations) to the Threlfall generators by $S=ab$ and $T=(ab)b^4(ab)$, and to the Cummins and Patera generators by $S=A_2A_1A_3A_2$ and $T=A_1A_2A_3$. \\ \vspace{-0.14in}

The generators $S$ and $T$ of the Shirai presentation have been given for the single-valued representations in \cite{shirai,A5}.  Rather than present the complete list for the double-valued representations, we provide the Shirai presentation generators for the doublets ${\bf 2}$ and ${\bf 2}^\prime$, which can also be found in \cite{shirai}:
\begin{eqnarray}\nonumber
S_2=\frac{1}{2}
\left(
\begin{array}{cc}
 i (\phi -1) & \phi +i \\
 i-\phi & i (1-\phi )
\end{array}
\right), \qquad 
T_2=\frac{1}{2}
\left(
\begin{array}{cc}
 \phi +i &  i (\phi -1) \\
  i (\phi -1) & \phi -i
\end{array}
\right)
\end{eqnarray}

\begin{eqnarray}\nonumber
S_{2^{\prime}}=\frac{1}{2}
\left(
\begin{array}{cc}
 i \phi  & \phi -(1-i) \\
(1+i)-\phi  & -i \phi 
\end{array}
\right), \qquad 
T_{2^{\prime}}=\frac{1}{2}
\left(
\begin{array}{cc}
 (1-i)-\phi  & -i \phi \\
 -i \phi  & (1+i)-\phi 
\end{array}
\right).
\end{eqnarray}
The Shirai presentation generators for the ${\bf 4}^\prime$ and the ${\bf 6}$  can also be found in \cite{shirai}.  Since we will not need them for the flavor model-building at leading order that we will consider in this paper, we do not state them here (we will consider more general scenarios in future work \cite{alexlisainprep}).  For completeness of presentation, however, we also state here the Shirai presentation generators  for the triplets ${\bf 3}$ and ${\bf 3}^\prime$, which are given by
\begin{eqnarray}
\label{stmat3}
S_{3}=\frac{1}{2}\left( \begin{array}{ccc}
 -1 &\phi& \frac{1}{\phi}\\
 \phi&\frac{1}{\phi}& 1\\
 \frac{1}{\phi}&1&-\phi
 \end{array} \right ), \;\;\;\;
T_{3}=\frac{1}{2}\left( \begin{array}{ccc}
 1&\phi&\frac{1}{\phi}\\
 -\phi&\frac{1}{\phi}& 1\\
\frac{1}{\phi}&-1&\phi
 \end{array} \right ),
\end{eqnarray}
\begin{eqnarray}
\label{stmat3p}
 S_{3^\prime}=\frac{1}{2}\left( 
\begin{array}{ccc}
 -\phi& \frac{1}{\phi}&1\\
 \frac{1}{\phi}&-1& \phi\\
 1&\phi&\frac{1}{\phi}
 \end{array} 
\right ),\;\;\;\;
T_{3^\prime}=\frac{1}{2}\left( \begin{array}{ccc}
 -\phi&-\frac{1}{\phi}&1\\
 \frac{1}{\phi}&1& \phi\\
- 1&\phi&-\frac{1}{\phi}
 \end{array} \right ).
\end{eqnarray}
While the group presentations of $\mathcal{I}$ and $\mathcal{I}^\prime$ are well known in the literature, the further group theoretical tools that are needed for flavor model building are the rules for constructing group invariants in a form that is suitable for physics applications.   For $\mathcal{I}$, we did this exercise for the Shirai presentation \cite{A5}.  Here we will present the basic relations for $\mathcal{I}^\prime$ that we have calculated in the Shirai presentation that are necessary for flavor model building at leading order, and defer the complete list of $\mathcal{I}^\prime$ tensor products for future work \cite{alexlisainprep}. \\ \vspace{-0.14in}

We begin with the tensor product ${\bf 2}\otimes {\bf 2} = {\bf 1}\oplus {\bf 3}$.  Defining the two distinct doublets as  ${\bf 2}=(a_1,a_2)^T$ and ${\bf 2}=(b_1,b_2)^T$, it is straightforward to show that up to normalization factors, the singlet is given by the antisymmetric combination
\begin{eqnarray}
{\bf 1}=a_2 b_1-a_1 b_2,
\label{singlet}
\end{eqnarray}
and the triplet ${\bf 3}$ is given by the symmetric combination
\begin{eqnarray}
{\bf 3}=( -i a_1 b_1+i a_2 b_2,  a_1 b_1+a_2 b_2,   i a_2 b_1+i a_1 b_2)^T.
\end{eqnarray}
For ${\bf 2}^\prime \otimes {\bf 2}^\prime = {\bf 1}\oplus {\bf 3}^\prime$, the singlet is again given by Eq.~(\ref{singlet}), and the triplet ${\bf 3}^\prime$ takes the (symmetric) form 
\begin{eqnarray}
{\bf 3}^\prime=(-i a_1 b_1-i a_2 b_2,  -a_1 b_1+a_2 b_2,  a_2 b_1+a_1 b_2)^T.
\end{eqnarray}
The tensor product of ${\bf 2} \otimes {\bf 2}^\prime={\bf 4}$ results in the following form for the quartic:
\begin{eqnarray}
{\bf 4}=(-a_2 b_1-a_1 b_2,  i a_2 b_1-i a_1 b_2, -i a_1 b_1-i a_2 b_2, -a_1 b_1+a_2 b_2)^T.
\end{eqnarray}
For ${\bf 2}\otimes {\bf 3}={\bf 2}\oplus{\bf 4}^\prime$, in which the initial doublet is $(a_1,a_2)^T$ and the initial triplet is $(b_1,b_2,b_3)^T$, the doublet is  
\begin{eqnarray}
{\bf 2}= (-a_2b_1+i a_2 b_2-a_1b_3, -a_1b_1-i a_1b_2+a_2 b_3)^T,
\end{eqnarray}
and the ${\bf 4}^\prime$ (which we will not need in this paper) takes the form
\begin{eqnarray}
{\bf 4}^\prime= \frac{1}{\sqrt{3}}(\sqrt{3}(i a_1b_1+a_1b_2), i a_2b_1+a_2b_2-2ia_1b_3,-i a_1b_1+a_1b_2-2ia_2b_3, \sqrt{3}(-ia_2b_1+a_2b_2))^T.
\end{eqnarray}
Similarly, for ${\bf 2}^\prime \otimes {\bf 3}^\prime={\bf 2}^\prime\oplus {\bf 4}^\prime$,  the ${\bf 2}^\prime$ is
\begin{eqnarray}
{\bf 2}^\prime= (ia_2b_1-a_2 b_2-a_1b_3, -ia_1b_1- a_1b_2+a_2 b_3)^T,
\end{eqnarray}
and the ${\bf 4}^\prime$ is given by 
\begin{eqnarray}
{\bf 4}^\prime &=& \frac{1}{\sqrt{5}}\left ( \begin{array}{c} -\frac{i}{\phi^2} a_2b_1-\phi^2 a_2b_2+\sqrt{5} a_1b_3 \\ \sqrt{3}(-i\phi a_1b_1+\frac{1}{\phi}a_1b_2-a_2b_3)\\ \sqrt{3}(-i\phi a_2b_1-\frac{1}{\phi}a_2b_2-a_1b_3)\\ 
 -\frac{i}{\phi^2} a_1b_1+\phi^2 a_ab_2+\sqrt{5} a_2b_3 \end{array} \right ),
\end{eqnarray}
in which $\phi=(1+\sqrt{5})/2$ is the golden ratio. Once again, for completeness, let us also recall that the singlet in the ${\bf 3}\otimes {\bf 3}={\bf 1}\oplus {\bf 3} \oplus {\bf 5}$ takes the form (as does the singlet in the analogous relation for ${\bf 3}^\prime \otimes {\bf 3}^\prime$):
\begin{eqnarray}
{\bf 1}=a_1b_1+a_2b_2+a_3 b_3,
\end{eqnarray}
in which we have written the initial triplets as $(a_1,a_2,a_3)^T$ and  $(b_1,b_2,b_3)^T$.  (For the explicit form of the ${\bf 3}$ and the ${\bf 5}$ and their primed counterparts, see \cite{A5}.)

\section{Quark Flavor Model Building and $U(2)$ Textures}

In this section, we turn to a discussion of quark flavor model building based on $\mathcal{I}^\prime$ and present a viable model of the quark masses and mixings based on $\mathcal{I}^\prime\otimes \mathcal{Z}_9$.  Let us begin by recalling the results of our previous study \cite{A5} of the icosahedral group $\mathcal{I}$ ($\mathcal{A}_5$), in which we constructed a simple toy lepton flavor model that resulted in a solar mixing angle given by $\tan\theta_{\rm sol}=1/\phi$.  In this analysis, we embedded the lepton doublets $L_i$ as a ${\bf 3}$ and the lepton singlets $e^c_i$ as a ${\bf 3}^\prime$, and constructed an effective neutrino seesaw matrix based on the relation ${\bf 3}\otimes {\bf 3}={\bf 1}\oplus {\bf 3}\oplus {\bf 5}$ and a charged lepton mass matrix based on ${\bf 3}\otimes {\bf 3}^\prime={\bf 4}\oplus {\bf 5}$.  To break the symmetry, we introduced a flavon field $\xi$ that transforms as a ${\bf 5}$ for the neutrino sector, and two charged lepton sector flavon fields $\chi$ and $\psi$, which transform as a ${\bf 4}$ and ${\bf 5}$, respectively.  The breaking of $\mathcal{I}$ by these flavon fields resulted in a toy model in which the solar mixing angle is obtained from the neutrino sector, the maximal atmospheric mixing angle is obtained from the charged leptons, and the reactor angle is zero at leading order.   Issues that were not addressed in \cite{A5} included the generation of the flavon field vacuum expectation values as well as higher order corrections; in forthcoming work, we will discuss these issues and construct other examples of lepton flavor models \cite{alexlisainprep}).   \\ \vspace{-0.14in}

 One result from our previous study is that it is not easy to generate the strongly hierarchical charged fermion masses within the $\mathcal{I}$ group.  More precisely, in our example the two flavon fields of the charged lepton sector had a delicate balance of vacuum expectation values in order to generate a structure in which only the $\tau$ lepton has a nonvanishing mass at leading order.  The challenge ultimately results in part because $\mathcal{I}$ only has one irreducible representation that is one-dimensional, which is the singlet representation ${\bf 1}$.  This is in sharp contrast to the case of $\mathcal{T}$ ($\mathcal{A}_4$), which has three one-dimensional representations: ${\bf 1}$, ${\bf 1}^\prime$, and ${\bf 1}^{\prime\prime}$.  The use of these three representations allows for the generation of mass hierarchies in the lepton sector through an additional $U(1)_{\rm FN}$ or other symmetries that result in specific higher-dimensional operators \cite{A4,ding}.   Even with this freedom in $\mathcal{A}_4$  models,  it is known that the quark sector is more easily accommodated by extending the group to $\mathcal{T}^\prime$, and embedding the quarks in doublet and one-dimensional representations \cite{carone,tprime,ding}.  Hence, we expect a similar extension is necessary for the icosahedral symmetry group, for which the generation of mass hierarchies is more difficult than that of the tetrahedral symmetry case.\\ \vspace{-0.14in}

For this reason, 
we now turn to the binary icosahedral group $\mathcal{I}^\prime$, and construct a supersymmetric model in which the quark superfields are embedded in doublet and singlet representations of $\mathcal{I}^\prime$.  
The standard approach is to assign the quark superfields of the third generation to singlet representations,
\begin{eqnarray}
Q_3 = (t,b)^T \rightarrow {\bf 1},\qquad t^c  \rightarrow {\bf 1},\qquad b^c  \rightarrow {\bf 1},
\label{thirdgen}
\end{eqnarray}
and assign the lighter generations to doublet representations.   Within $\mathcal{I}^\prime$, there are two distinct doublets, ${\bf 2}$ and ${\bf 2}^\prime$, which allows for some flexibility in model-building.  In this paper, we will assign the first and second generation quark superfields to the ${\bf 2}$ representation, as follows:
\begin{eqnarray}
Q = (Q_1,Q_2)^T = ((u,d),\, (c,s))^T\rightarrow {\bf 2},\qquad u^c=(u^c,c^c)^T  \rightarrow {\bf 2},\qquad d^c =(d^c, s^c) \rightarrow {\bf 2}.
\label{firstsecondgen}
\end{eqnarray}
This assignment will allow us to construct models with the $U(2)$ relation, ${\bf 2}\otimes {\bf 2}={\bf 1}\oplus {\bf 3}$.  We note that a replacement of ${\bf 2}$ by ${\bf 2}^\prime$ in Eq.~(\ref{firstsecondgen}) would also result in an analogous relation, ${\bf 2}^\prime \otimes {\bf 2}^\prime={\bf 1}\oplus {\bf 3}^\prime$, which would result in a similar quark flavor models (at least at leading order).   An alternative assignment of $Q$ to the ${\bf 2}$ and $u^c$, $d^c$ to the ${\bf 2}^\prime$ representations (or permutations) would yield models based on the relation ${\bf 2}\otimes {\bf 2}^\prime ={\bf 4}$, which would necessitate a different pattern of flavor symmetry breaking. We defer the investigation of this possibility to forthcoming work \cite{alexlisainprep}.\\ \vspace{-0.14in}

To break the family symmetry, we first recall that in $U(2)$ flavor models with this assignment of the Standard Model quarks, the family symmetry is broken in two stages: first $U(2)$ is broken to $U(1)$ by nonvanishing vacuum expectation values of triplet and doublet fields, and next the $U(1)$ is broken by fields that are singlets under the original family symmetry.   This basic flavon sector is retained in $U(2)$-inspired models based on discrete non-Abelian family symmetries such as $\mathcal{T}^\prime$ (see for example \cite{ding}), and we will also use it here.  
Hence, we now introduce a flavon chiral superfield $\psi$ that transforms as a ${\bf 3}$ and  two flavon chiral superfields $\eta_{1,2}$ that each transform as a ${\bf 2}$ under $\mathcal{I}^\prime$, as well as a set of $\mathcal{I}^\prime$ singlet superfields $\rho$, $\sigma$, and $\chi$.  \\ \vspace{-0.14in}

\begin{table}[tbp]
\begin{tabular}{|c||c|c|c|c|c|c|c||c|c|c|c|c|c||c|c|c|c|}\hline
Field&$Q$&$Q_3$&$u^c$&$d^c$&$t^c$&$b^c$&$H_{u,d}$&$\rho$&$\sigma$&$\chi$&$\eta$&$\eta_2$&$\psi$& $\sigma^0$ &$\chi^0$& $\eta^0$ & $\psi^0$ \\ \hline 
$\mathcal{I}^{\prime}$&2&1&2&2&1&1&1&1&1&1&2&2&3&1&1&2&3\\\hline
$\mathcal{Z}_9$&1&$\alpha$ &$\alpha^8$&$\alpha^5$&$\alpha^8$&$\alpha^6$&1&$\alpha^8$&$\alpha^5$&$\alpha^2$&1&$\alpha$&$\alpha^8$ &$\alpha^5$ & $\alpha^2$& 1 &$\alpha^8$ \\\hline
\end{tabular}
\vspace{0.1in}
\caption{\label{t1}
The charge assignments for the quark, Higgs, flavon, and driving field supermultiplets in our $\mathcal{I}^\prime\otimes \mathcal{Z}_9$ model, with $\alpha=e^{2\pi i/9}$.  The flavon fields are the $\rho$, $\sigma$, $\chi$, $\eta$, $\eta_2$, and $\psi$, while the driving fields are $\sigma^0$, $\chi^0$, $\eta^0$, and $\psi^0$.}
\end{table}
To obtain realistic quark mass matrices, it is necessary to augment $\mathcal{I}$ with an additional symmetry, which we take to be a $\mathcal{Z}_9$ group.   The $\mathcal{I}^\prime\otimes \mathcal{Z}_9$ charge assignments for the SM matter superfields and flavon sector superfields of our model in presented in Table~\ref{t1}.  We have assumed that the MSSM Higgs doublets $H_{u,d}$ are inert with respect to the $\mathcal{I}^\prime\otimes \mathcal{Z}_9$ symmetry.
We note that the choice of $\mathcal{Z}_9$ as an additional symmetry of the quark sector has also been made in the $\mathcal{T}^\prime \otimes \mathcal{Z}_3\otimes \mathcal{Z}_9$ model of \cite{ding}, though our flavon field content and charge assignments are different.  With these charge assignments, the superpotential  terms that result in effective Yukawa terms involving the up-type quarks take the following form:
\begin{eqnarray}
W_u&=&y_{u1} Q_3 t^c H_u+\frac{y_{u2}}{M} Q_3 u^c \eta_1 H_u +\frac{y_{u2}^\prime}{M^2} Q_3 u^c \eta_2 \rho H_u +\frac{y_{u2}^{\prime\prime}}{M^2} Q_3 u^c \eta_2 \psi H_u+\frac{y_{u3}}{M}Q t^c \eta_2H_u\nonumber \\&+&\frac{y_{u4}}{M^2}Q u^c \sigma \sigma H_u+\frac{y_{u4}^\prime}{M^2}Q u^c \rho \chi H_u+\frac{y_{u4}^{\prime\prime}}{M^2}Q u^c \eta \eta_2 H_u+\frac{y_{u4}^{\prime\prime\prime}}{M^2}Q u^c  \chi \psi H_u,
\label{ups}
\end{eqnarray}
in which $M$ represents the (presumably high) cutoff scale of the effective theory, and the $y_{ui}$ are dimensionless (order one) couplings.  Similarly, the effective Yukawa couplings for the down-type quark superfields are given by
\begin{eqnarray}
W_d&=&\frac{y_{d1}}{M} Q_3 b^c \chi H_d+\frac{y_{d2}}{M^2}Q_3d^c \eta_2\chi H_d+\frac{y_{d3}}{M^2}Q b^c \eta_2\chi H_d+\frac{y_{d4}}{M^2}Q d^c\rho \sigma H_d+\frac{y_{d4}^\prime}{M^2}Q d^c\chi \chi H_d+\frac{y_{d4}^{\prime\prime}}{M^2}Q d^c \sigma \psi H_d,
\label{downs}
\end{eqnarray}
in which the $y_{di}$ again represent dimensionless couplings. To break the flavor symmetry, we assume that the flavon fields develop vacuum expectation values as follows (this form will be justified later in the paper):
\begin{eqnarray}
\langle \psi \rangle &=& \langle (\psi^1,\psi^2,\psi^3)^T\rangle =\frac{v_3}{2}(-i,1,0)^T, \;\; \langle \eta_1 \rangle =\langle (\eta_1^1,\eta_1^2)^T\rangle =(v_{21},0)^T,\nonumber \\    \langle \eta_2 \rangle &=&\langle (\eta_2^1,\eta_2^2)^T\rangle =(v_{22},0)^T,\;\;
\langle \rho \rangle = v_\rho, \;\; \langle \chi \rangle = v_\chi,\;\; \langle \sigma \rangle = v_\sigma,
\label{vevs}
\end{eqnarray}
in which the vacuum expectation values are of the order
\begin{eqnarray}
\left \vert \frac{v_3}{M} \right \vert \sim \left \vert \frac{v_{21}}{M} \right \vert \sim \left \vert \frac{v_{22}}{M} \right \vert  \sim \lambda^2, \qquad
\left \vert \frac{v_\rho}{M} \right \vert \sim \left \vert \frac{v_\chi}{M} \right \vert \sim \left \vert \frac{v_\sigma}{M} \right \vert  \sim \lambda^3,
\label{vevvalues}
\end{eqnarray}
where $\lambda\equiv \sin\theta_c=0.22$ is the Cabibbo angle.  Upon flavor and electroweak symmetry breaking (with $\langle H_{u,d}\rangle=v_{u,d}$), the quark mass matrices that result from Eq.~(\ref{ups}) and Eq.~(\ref{downs}) take the form
\begin{eqnarray}
\mathcal{M}_u=\left ( \begin{array}{ccc}  \vspace{0.08in} 0&-y_{u4}\frac{v_\sigma^2}{M^2}-y_{u4}^\prime\frac{v_\rho v_\chi}{M^2}&0\\ \vspace{0.08in}
y_{u4}\frac{v_\sigma^2}{M^2}+y_{u4}^\prime\frac{v_\rho v_\chi}{M^2}&y_{u4}^{\prime\prime}\frac{v_{21}v_{22}}{M^2}+y_{u4}^{\prime\prime\prime}\frac{v_3v_\chi}{M^2}&y_{u3}\frac{v_{22}}{M}\\ 0& y_{u2}\frac{v_{21}}{M}+y_{u2}^\prime \frac{v_{22}v_\rho}{M^2}& y_{u1}   \end{array}\right )v_u \ \equiv  \ \left (\begin{array}{ccc} \vspace{0.08in} 0& -\tilde{y}_{u4}\lambda^6 & 0\\ \vspace{0.08in} \tilde{y}_{u4}\lambda^6 & \tilde{y}_{u4}^{\prime\prime}\lambda^4+\tilde{y}_{u4}^{\prime\prime\prime}\lambda^5  & \tilde{y}_{u3}\lambda^2 \\ 0 & \tilde{y}_{u2}\lambda^2 & \tilde{y}_{u1} \end{array} \right )v_u,
\end{eqnarray}
and
\begin{eqnarray}
 \mathcal{M}_d=\left ( \begin{array}{ccc}  \vspace{0.08in} 0& -y_{d4} \frac{v_\rho v_\sigma}{M^2}-y_{d4}^\prime \frac{v_\chi^2}{M^2}&0\\ \vspace{0.08in}
 y_{d4} \frac{v_\rho v_\sigma}{M^2}+y_{d4}^\prime \frac{v_\chi^2}{M^2}&y_{d4}^{\prime\prime} \frac{v_3v_\sigma}{M^2}&
 y_{d3}\frac{v_{22}v_\chi}{M^2}\\ 0& y_{d2}\frac{v_{22}v_\chi}{M^2}& y_{d1}\frac{v_\chi}{M} \end{array}\right )v_d  \ \equiv  \ \left (\begin{array}{ccc} \vspace{0.08in} 0& -\tilde{y}_{d4}\lambda^3 & 0\\ \vspace{0.08in} \tilde{y}_{d4}\lambda^3 & \tilde{y}_{d4}^{\prime\prime}\lambda^2 & \tilde{y}_{d3}\lambda^2 \\ 0 & \tilde{y}_{d2}\lambda^2 & \tilde{y}_{d1} \end{array} \right ) \lambda^3 v_d.
\end{eqnarray}
It is straightforward to diagonalize these mass matrices using perturbation theory, which yields the following results for the quark masses at leading order:
\begin{eqnarray}
\label{masses}
m_u &\simeq&  \frac{\vert \tilde{y}_{u1}\vert \vert \tilde{y}_{u4}\vert^2}{\vert \tilde{y}_{u2} \tilde{y}_{u3}-\tilde{y}_{u1}\tilde{y}_{u4}\vert} \lambda^8 v_u, 
\qquad m_c \simeq \left \vert \frac{\tilde{y}_{u2}\tilde{y}_{u3}}{\tilde{y}_{u1}}-\tilde{y}_{u4}^{\prime\prime}\right \vert \lambda^4 v_u, 
\qquad m_t \simeq \left (\vert \tilde{y}_{u1}\vert +\frac{\vert \tilde{y}_{u2} \vert^2+\vert \tilde{y}_{u3} \vert^2}{2 \vert \tilde{y}_{u1}\vert }\lambda^4\right ) v_u,\\
m_d&\simeq & \frac{\vert \tilde{y}_{d4} \vert^2}{\vert \tilde{y}_{d4}^{\prime\prime}\vert}\lambda^7v_d, \qquad \qquad \qquad \;\;\;\;
m_s\simeq \vert \tilde{y}_{d4}^{\prime\prime} \vert \lambda^5 v_d,
\qquad \qquad \qquad \;\; m_b\simeq \left (\vert \tilde{y}_{d1}\vert + \frac{\vert \tilde{y}_{d2} \vert^2+\vert \tilde{y}_{d3} \vert^2}{2\vert \tilde{y}_{d1}\vert }\lambda^4 \right )\lambda^3v_d.\nonumber
\end{eqnarray}
Hence, the quark mass ratios are predicted have the appropriate powers of the Cabibbo angle $\lambda$, which are given as follows:
: $m_u:m_c:m_t\sim \lambda^8:\lambda^4:1$ and $m_d:m_s:m_b\sim \lambda^4:\lambda^2:1$.  We also have $m_b:m_t \sim \lambda^3:1$, which is consistent with low to moderate values of $\tan\beta=v_u/v_d$, as in the $\mathcal{T}^\prime$ model of \cite{ding}. This range of $\tan\beta$ has advantages in terms of model-building compared with larger values of $\tan\beta$, such as the avoidance of large radiative corrections.\\ \vspace{-0.14in}

Similarly, the leading order elements of the CKM mixing matrix also have the appropriate powers of $\lambda$:
\begin{eqnarray}
V_{ud} &\sim&  V_{cs}\sim  V_{tb} \sim 1,\;\;\;  V_{us}  \sim -\frac{\tilde{y}_{d4}}{\tilde{y}_{d4}^{\prime\prime }}\lambda- \frac{\tilde{y}_{u1}\tilde{y}_{u4}}{\tilde{y}_{u2}\tilde{y}_{u3}-\tilde{y}_{u1}\tilde{y}_{u4}^{\prime\prime}}\lambda^2 \sim -V_{cd}^*,
 \;\;\;
V_{cb}\sim \left (\frac{\tilde{y}_{d3}}{\tilde{y}_{d1}}-\frac{\tilde{y}_{u3}}{\tilde{y}_{u1}}\right )\lambda^2 \sim -V_{ts}^*,
 \nonumber \\
 V_{ub}&\sim& \frac{\tilde{y}_{u4}\tilde{y}_{u1}}{\tilde{y}_{u2}\tilde{y}_{u3}-\tilde{y}_{u1}\tilde{y}_{u4}^{\prime\prime}}\left (\frac{\tilde{y}_{u3}}{\tilde{y}_{u1}}-\frac{\tilde{y}_{d3}}{\tilde{y}_{d1}} \right )\lambda^4, \;\;
 \qquad  V_{td} \sim \frac{\tilde{y}_{d4}^*}{\tilde{y}_{d4}^{\prime\prime \,*}} \left (-\frac{\tilde{y}_{d3}^*}{\tilde{y}_{d1}^*}+\frac{\tilde{y}_{u3}^*}{\tilde{y}_{u1}^*} \right )\lambda^3.
\label{ckm}
\end{eqnarray}
From Eqs.~(\ref{masses})--(\ref{ckm}), we see that the well-known $U(2)$ relations $\vert V_{td}/V_{ts} \vert = \sqrt{m_d/m_s}$ and  $\vert V_{ub}/V_{cb} \vert = \sqrt{m_u/m_c}$ are reproduced at leading order, as is the case in the $\mathcal{T}^\prime$ model of \cite{ding}.  
\\ \vspace{-0.14in}  

Therefore, we see that the embedding of the quarks as given in Eq.~(\ref{thirdgen}) and Eq.~(\ref{firstsecondgen}) within $\mathcal{I}^\prime$ can result in a viable quark flavor model at leading order, provided the assumption of the specific flavon sector as given in Table~\ref{t1}.   In principle, we can go further and compute higher order corrections to the results of Eq.~(\ref{masses}) and Eq.~(\ref{ckm}), which will modify the leading order relations.    However, we do not do so in this paper because we are not including the leptons, and the flavons needed to break the family symmetry might contribute in the quark sector at higher order.  We comment that in the lepton model we presented in \cite{A5}, the lepton sector flavons are fields that transform either as a ${\bf 4}$ or a ${\bf 5}$ of $\mathcal{I}$, and so these fields will not contribute at leading order to the quark mass matrices of the $U(2)$-inspired model presented here simply due to icosahedral symmetry.  Clearly, for alternate embeddings in which the quarks of the lighter generations are embedded in both the ${\bf 2}$ and ${\bf 2}^\prime$, any lepton sector flavon field that transforms as a ${\bf 4}$ of $\mathcal{I}$ is allowed by icosahedral symmetry to couple to the quarks at leading order, though additional discrete symmetries can also be imposed to forbid such couplings. \\ \vspace{-0.14in}

We now turn to the important question of justifying the family symmetry breaking pattern of Eq.~(\ref{vevs}).  To address the dynamics of the flavon sector, we follow the standard approach and recall that a global $U(1)_R$ charge is present in the supersymmetric sector of the theory (it is broken to a discrete R-parity when supersymmetry breaking terms are included), such that the superpotential terms satisfy the constraint that the total $R$-charge of any allowed term is $+2$.  Hence, it is necessary to introduce additional ``driving" fields which have a $U(1)_R$ charge of $+2$  and thus couple linearly to the flavon fields in the superpotential.   Our choice of driving fields is presented in Table~\ref{t1}.  These fields include two $\mathcal{I}^\prime$ singlets $\sigma^0$ and $\chi^0$, one $\mathcal{I}^\prime$ doublet $\eta^0$, and one  $\mathcal{I}^\prime$ triplet $\psi^0$.  With these charge assignments, the leading order superpotential couplings involving the flavon fields and the driving fields take the form
\begin{eqnarray}
W_{\rm fl}=M_\eta \eta_1 \eta^0+g_1\eta^0\eta_2 \rho+g_2 \sigma^0 \sigma \rho+g_3 \sigma^0\chi \chi+g_4 \chi^0 \rho \rho+g_5 \chi^0 \sigma \chi+g_6 \chi^0 \psi \psi+g_7 \eta^0 \eta_2 \psi+g_8  \eta_1 \eta_2 \psi^0+g_9 \chi \psi \psi^0.
\end{eqnarray}
With the assumption that the soft supersymmetry breaking scalar mass-squared parameters of the driving fields are all positive at the symmetry breaking scale, the driving fields do not develop vacuum expectation values, and hence it is only necessary to enforce that the F terms of the driving fields vanish to obtain a local minimum of the flavon field potential.  These F terms are given as follows:
\begin{eqnarray}
\label{fterms}
\frac{\partial W_{\rm fl}}{\partial \sigma^0}&=& g_2 \sigma \rho +g_3 \chi\chi,\qquad \frac{\partial W_{\rm fl}}{\partial \chi^0}= g_4 \rho \rho+g_5 \sigma \chi +g_6 (\psi^1 \psi^1+\psi^2 \psi^2+\psi^3 \psi^3),\\
\frac{\partial W_{\rm fl}}{\partial \eta^{0\,1}}&=&M_\eta \eta_1^{2}+g_7 \eta_2^{1}(-i\psi^1+\psi^2)+g_7 \eta_2^2 i \psi^3+g_1\rho \eta_2^2,\;\;  
\frac{\partial W_{\rm fl}}{\partial \eta^{0\,2}}=
-M_\eta \eta_1^{1}+g_7 \eta_2^{2}(i\psi^1+\psi^2)+g_7 \eta_2^1 i \psi^3-g_1\rho \eta_2^1,\nonumber \\
\frac{\partial W_{\rm fl}}{\partial \psi^{0\,1}}&=& g_8 (-i \eta_1^1 \eta_2^1+i \eta_1^2\eta_2^2)+g_9 \chi \psi^1,\;\;
\frac{\partial W_{\rm fl}}{\partial \psi^{0\,2}} = g_8 (\eta_1^1 \eta_2^1+\eta_1^2\eta_2^2)+g_9 \chi \psi^2, \;\; 
\frac{\partial W_{\rm fl}}{\partial \psi^{0\,3}} 
=g_8 (i\eta_1^2 \eta_2^1+i\eta_1^1\eta_2^2)+g_9 \chi \psi^3.\nonumber
\end{eqnarray}
The requirement that the F terms of Eq.~(\ref{fterms}) vanish at the minimum is consistent with the form of the flavon vacuum expectation values of Eq.~(\ref{vevs}), in which the vevs satisfy the following relations:
\begin{eqnarray}
v_\sigma=-\frac{g_4}{g_5}\left (\frac{g_3g_5}{g_4g_2} \right )^{1/3}v_\rho, \;\;\;\; v_\chi=v_\rho \left (\frac{g_2g_4}{g_3g_5} \right )^{1/3}, \;\;\;\; v_{21}=-\frac{g_1 v_{22}v_\rho}{M_\eta},\;\;\;\; v_3= \frac{2g_1 v_{22}^2}{M_\eta}\frac{g_8}{g_9}\left (\frac{g_3g_5}{g_4g_2} \right )^{1/3},
\end{eqnarray}
in which the flat directions $v_\rho$ and $v_{22}$ are presumably lifted by supersymmetry breaking terms.
For order one values of the coupling ratios $g_4/g_5$ and $g_2/g_3$, the singlet vacuum expectation values are naturally $v_\rho\sim v_\sigma \sim v_\chi$.  For the doublet and triplet flavons, a tuning of $M_\eta/g_1 \sim \lambda^{3}M$ and $g_8/g_9\sim \lambda$ is required to obtain $v_{21}\sim v_{22}\sim v_3$ with order one values of $g_4/g_5$ and $g_2/g_3$.   Hence, the vacuum structure specified by Eq.~(\ref{vevs}) and Eq.~(\ref{vevvalues}) can be obtained within a specific subset of the multidimensional parameter space of the flavon sector potential.

\section{Conclusions}
With the advent of the large angles of the lepton mixing data, the paradigm of discrete non-Abelian family symmetries has become standard in efforts to address the fermion mass puzzle of the Standard Model.  Of the possible discrete non-Abelian symmetry groups to consider as a basis for flavor model building, we have argued in this paper that the double cover of the rotational icosahedral symmetry group, the binary icosahedral group $\mathcal{I}^\prime$, is well-suited for this purpose due to its nontrivial triplet and doublet representations.  In comparison to other discrete groups based on the Platonic solids, there has been a relative paucity of appearances of $\mathcal{I}^\prime$ within the physics literature, which is due at least in part to the fact that the icosahedron does not obey translational symmetry.  Therefore, we have presented the basic group theory of $\mathcal{I}^\prime$, which is available primarily in the mathematics literature, in a form that is suitable for flavor model building.  As an example, we applied $\mathcal{I}^\prime$ as a family symmetry of the quark sector and  constructed a viable $\mathcal{I}^\prime \otimes \mathcal{Z}_9$ quark flavor model that reproduces many of the desirable features of the well-known $U(2)$ flavor models and other models based on the fundamental relation ${\bf 2}\otimes {\bf 2}={\bf 1}\oplus {\bf 3}$.  By analyzing the dynamics of the flavon fields that break the family symmetry,  we have demonstrated that the successful leading order predictions of this model can be accommodated within a subset of the full parameter space of the theory.\\ \vspace{-0.14in}

With respect to other discrete non-Abelian family symmetry groups that have been used for flavor model building, the icosahedral group and its double cover are relatively large groups (with group orders of 60 and 120, respectively) which have larger representations.  Therefore, icosahedral symmetry provides a very rich setting in which to address the fermion mass puzzle of the Standard Model.   Additional studies are needed to determine the full power of $\mathcal{I}^\prime$ as a viable family symmetry group for the quark and lepton sectors.  We defer further investigations of this intriguing framework to future work.

\acknowledgments
We thank D.~J.~H.~Chung for helpful conversations and comments. This work was supported by the University of Wisconsin Alumni Research Foundation and the U. S. Department of Energy contract DE-FG-02-95ER40896.



\begin{thebibliography}{}
\bibitem{SK}
	Y. Fukuda {\it et al.}, [Super-Kamiokande Collaboration], \Journal{\PRL}{81}{1562}{1998}; \Journal{\PRL}{82}{2430}{1999};
	T. Kajita for the collaboration, \Journal{\NPSUPPL}{77}{123}{1999}.
	See also
	T. Kajita and Y. Totsuka, \Journal{\RMP}{73}{85}{2001}.

\bibitem{K2K}
	S. H. Ahn, {\it et al.}, [K2K Collaboration], \Journal{\PLB}{511}{178}{2001}; \Journal{\PRL}{90}{041801}{2003}.

\bibitem{OldSolar}
    J.N. Bahcall, W.A. Fowler, I. Iben and R.L. Sears, \Journal{\APJ}{137}{344}{1963};
    J. Bahcall, \Journal{\PRL}{12}{300}{1964};
    R. Davis, Jr., \Journal{\PRL}{12}{303}{1964};
    R. Davis, Jr., D.S. Harmer and K.C. Hoffman, \Journal{\PRL}{20}{1205}{1968}; 
    J.N. Bahcall, N.A. Bahcall and G. Shaviv, \Journal{\PRL}{20}{1209}{1968};
    J.N. Bahcall and R. Davis, Jr., \Journal{\SCI}{191}{264}{1976}.

\bibitem{Sun}
	Y. Fukuda {\it et al.}, [Super-Kamiokande Collaboration], \Journal{\PRL}{81}{1158}{1998}; [\Journal{\Erratum}{81}{4279}{1998}];
	B.T. Clevel {\it et al.}, \Journal{\APJ}{496}{505}{1998};
	W. Hampel {\it et al.}, [GNO Collaboration],  \Journal{\PLB}{447}{127}{1999};
	Q.A. Ahmed. {\it et al.}, [SNO Collaboration], \Journal{\PRL}{87}{071301}{2001}; \Journal{\PRL}{89}{011301}{2002}.

\bibitem{Reactor} 
	K. Eguchi, {\it et al.}, [KamLAND collaboration], \Journal{\PRL}{90}{021802}{2003}.
	K. Inoue, [KamLAND collaboration], \Journal{\NJP}{6}{147}{2004};
	S. Abe et al., [KamLAND collaboration],  \Journal{\PRL}{100}{221803}{2008}.

\bibitem{Theta13}
	M. Apollonio, {\it et al.}, [CHOOZ Collaboration], \Journal{\EPJ}{27}{331}{2003}.
	
\bibitem{NuData}
 T. Schwetz, M. Tortola and J.W.F. Valle, \Journal{\NJP}{10}{113011}{2008}.
	

\bibitem{PMNS} 
	B. Pontecorvo, \Journal{\JETPUSSR}{7}{172}{1958} [\Journal{\ZETP}{34}{247}{1958}];
	Z. Maki, M. Nakagawa and S. Sakata, \Journal{\PTP}{28}{870}{1962}. 

\bibitem{kl}
Z.G.Berezhiani and M.Yu.Khlopov,
Yadernaya Fizika (1990) V. 51, 1157-1170
[English translation: Sov.J.Nucl.Phys. (1990) V. 51, 739-746];
Z.G.Berezhiani and M.Yu.Khlopov,
Yadernaya Fizika (1990) V. 51, 1479-1491
[English translation: Sov.J.Nucl.Phys. (1990) V. 51, PP. 935-942];
Z.G.Berezhiani and M.Yu.Khlopov,
Yadernaya Fizika (1990) V.~52,  96-103 [English translation:
Sov.J.Nucl.Phys. (1990) V.~52,  60-64];
Z.G.Berezhiani, M.Yu.Khlopov, and R.R.Khomeriki,
Yadernaya Fizika (1990) V.~52, 538-543 [English
translation: Sov.J.Nucl.Phys. (1990) V.52, 344-347];
Z.G.Berezhiani and M.Yu.Khlopov,
Z.Phys.C- Particles and Fields (1991), V.~49, 73-78;
A.S.Sakharov and M.Yu.Khlopov,
Yadernaya Fizika (1994) V.~57, 690-697. [English translation:
Phys.Atom.Nucl. (1994) V.~57, 651-658].


\bibitem{FN}	
  C.D. Froggatt and H.B. Nielsen, Nucl. Phys. B 147 (1979) 277.
	







\bibitem{A4}
  E.~Ma and G.~Rajasekaran,
  Phys.\ Rev.\  D {\bf 64}, 113012 (2001)
  [arXiv:hep-ph/0106291];
  K.~S.~Babu, E.~Ma and J.~W.~F.~Valle,
  Phys.\ Lett.\  B {\bf 552}, 207 (2003)
  [arXiv:hep-ph/0206292]; 
  E.~Ma,
  Phys.\ Rev.\  D {\bf 70}, 031901 (2004)
  [arXiv:hep-ph/0404199];
  G.~Altarelli and F.~Feruglio,
  Nucl.\ Phys.\  B {\bf 720}, 64 (2005)
  [arXiv:hep-ph/0504165];
  E.~Ma,
  Phys.\ Rev.\  D {\bf 72}, 037301 (2005)
  [arXiv:hep-ph/0505209]; 
  A.~Zee,
  Phys.\ Lett.\  B {\bf 630}, 58 (2005)
  [arXiv:hep-ph/0508278];
  G.~Altarelli and F.~Feruglio,
  Nucl.\ Phys.\  B {\bf 741}, 215 (2006)
  [arXiv:hep-ph/0512103];
  E.~Ma,
  Mod.\ Phys.\ Lett.\  A {\bf 21}, 2931 (2006)
  [arXiv:hep-ph/0607190]; 
  S.~F.~King and M.~Malinsky,
  Phys.\ Lett.\  B {\bf 645}, 351 (2007)
  [arXiv:hep-ph/0610250]; 
  S.~Morisi, M.~Picariello and E.~Torrente-Lujan,
  Phys.\ Rev.\  D {\bf 75}, 075015 (2007)
  [arXiv:hep-ph/0702034];
  M.~Honda and M.~Tanimoto,
  Prog.\ Theor.\ Phys.\  {\bf 119}, 583 (2008)
  [arXiv:0801.0181 [hep-ph]];
  G.~Altarelli, F.~Feruglio and C.~Hagedorn,
  JHEP {\bf 0803}, 052 (2008)
  [arXiv:0802.0090 [hep-ph]];
  P.~H.~Frampton and S.~Matsuzaki,
  arXiv:0806.4592 [hep-ph];
  F.~Feruglio, C.~Hagedorn, Y.~Lin and L.~Merlo,
  Nucl.\ Phys.\  B {\bf 809}, 218 (2009)
  [arXiv:0807.3160 [hep-ph]];
  P.~Ciafaloni, M.~Picariello, E.~Torrente-Lujan and A.~Urbano,
  Phys.\ Rev.\  D {\bf 79}, 116010 (2009)
  [arXiv:0901.2236 [hep-ph]];
  C.~Hagedorn, E.~Molinaro and S.~T.~Petcov,
  JHEP {\bf 0909}, 115 (2009)
  [arXiv:0908.0240 [hep-ph]];
  F.~Feruglio, C.~Hagedorn and L.~Merlo,
  JHEP {\bf 1003}, 084 (2010)
  [arXiv:0910.4058 [hep-ph]];
  J.~Berger and Y.~Grossman,
  JHEP {\bf 1002}, 071 (2010)
  [arXiv:0910.4392 [hep-ph]];
  F.~Feruglio, C.~Hagedorn, Y.~Lin and L.~Merlo,
  arXiv:0911.3874 [hep-ph];
  I.~K.~Cooper, S.~F.~King and C.~Luhn,
  Phys.\ Lett.\  B {\bf 690}, 396 (2010)
  [arXiv:1004.3243 [hep-ph]];
  A.~Kadosh and E.~Pallante,
  JHEP {\bf 1008}, 115 (2010)
  [arXiv:1004.0321 [hep-ph]].



\bibitem{carone}
  A.~Aranda, C.~D.~Carone and R.~F.~Lebed,
  Phys.\ Lett.\  B {\bf 474}, 170 (2000)
  [arXiv:hep-ph/9910392];
  A.~Aranda, C.~D.~Carone and R.~F.~Lebed,
  Int.\ J.\ Mod.\ Phys.\  A {\bf 16S1C}, 896 (2001)
  [arXiv:hep-ph/0010144];
  A.~Aranda, C.~D.~Carone and R.~F.~Lebed,
  Phys.\ Rev.\  D {\bf 62}, 016009 (2000)
  [arXiv:hep-ph/0002044].


\bibitem{tprime}
 F.~Feruglio, C.~Hagedorn, Y.~Lin and L.~Merlo,
  Nucl.\ Phys.\  B {\bf 775}, 120 (2007)
  [Erratum-ibid.\  {\bf 836}, 127 (2010)]
  [arXiv:hep-ph/0702194];
  M.~C.~Chen and K.~T.~Mahanthappa,
  Phys.\ Lett.\  B {\bf 652}, 34 (2007)
  [arXiv:0705.0714 [hep-ph]]; 
  A.~Aranda,
  Phys.\ Rev.\  D {\bf 76}, 111301 (2007)
  [arXiv:0707.3661 [hep-ph]];
  P.~H.~Frampton and T.~W.~Kephart,
  JHEP {\bf 0709}, 110 (2007)
  [arXiv:0706.1186 [hep-ph]];
  M.~C.~Chen and K.~T.~Mahanthappa,
  arXiv:0710.2118 [hep-ph];
  S.~Sen,
  Phys.\ Rev.\  D {\bf 76}, 115020 (2007)
  [arXiv:0710.2734 [hep-ph]];
  P.~H.~Frampton, T.~W.~Kephart and S.~Matsuzaki,
  Phys.\ Rev.\  D {\bf 78}, 073004 (2008)
  [arXiv:0807.4713 [hep-ph]];
  M.~C.~Chen and K.~T.~Mahanthappa,
  Phys.\ Lett.\  B {\bf 681}, 444 (2009)
  [arXiv:0904.1721 [hep-ph]];
  M.~C.~Chen, K.~T.~Mahanthappa and F.~Yu,
  Phys.\ Rev.\  D {\bf 81}, 036004 (2010)
  [arXiv:0907.3963 [hep-ph]];
  L.~Merlo,
  arXiv:1004.2211 [hep-ph].


\bibitem{ding}
  G.~J.~Ding,
  Phys.\ Rev.\  D {\bf 78}, 036011 (2008)
  [arXiv:0803.2278 [hep-ph]].


\bibitem{delta}
  E.~Ma,
  Mod.\ Phys.\ Lett.\  A {\bf 21}, 1917 (2006)
  [arXiv:hep-ph/0607056]; 
  I.~de Medeiros Varzielas, S.~F.~King and G.~G.~Ross,
  Phys.\ Lett.\  B {\bf 648}, 201 (2007)
  [arXiv:hep-ph/0607045];
  C.~Luhn, S.~Nasri and P.~Ramond,
  J.\ Math.\ Phys.\  {\bf 48}, 073501 (2007)
  [arXiv:hep-th/0701188];
  E.~Ma,
  Phys.\ Lett.\  B {\bf 660}, 505 (2008)
  [arXiv:0709.0507 [hep-ph]];
  J.~A.~Escobar and C.~Luhn,
  J.\ Math.\ Phys.\  {\bf 50}, 013524 (2009)
  [arXiv:0809.0639 [hep-th]];

\bibitem{Z3xZ7}
  C.~Luhn, S.~Nasri and P.~Ramond,
  Phys.\ Lett.\  B {\bf 652}, 27 (2007)
  [arXiv:0706.2341 [hep-ph]];
  C.~Hagedorn, M.~A.~Schmidt and A.~Y.~Smirnov,
  Phys.\ Rev.\  D {\bf 79}, 036002 (2009)
  [arXiv:0811.2955 [hep-ph]].

\bibitem{raidal}
  Y.~Kajiyama, M.~Raidal and A.~Strumia,
  Phys.\ Rev.\  D {\bf 76}, 117301 (2007)
  [arXiv:0705.4559 [hep-ph]].




\bibitem{S4}
  E.~Ma,
  Phys.\ Lett.\  B {\bf 632}, 352 (2006)
  [arXiv:hep-ph/0508231];
  C.~Hagedorn, M.~Lindner and R.~N.~Mohapatra,
  JHEP {\bf 0606}, 042 (2006)
  [arXiv:hep-ph/0602244];
  Y.~Cai and H.~B.~Yu,
  Phys.\ Rev.\  D {\bf 74}, 115005 (2006)
  [arXiv:hep-ph/0608022];
  H.~Zhang,
  Phys.\ Lett.\  B {\bf 655}, 132 (2007)
  [arXiv:hep-ph/0612214];
  Y.~Koide,
  JHEP {\bf 0708}, 086 (2007)
  [arXiv:0705.2275 [hep-ph]];
  F.~Bazzocchi, L.~Merlo and S.~Morisi,
  Nucl.\ Phys.\  B {\bf 816}, 204 (2009)
  [arXiv:0901.2086 [hep-ph]].
  Y.~Daikoku and H.~Okada,
  Phys.\ Rev.\  D {\bf 82}, 033007 (2010)
  [arXiv:0910.3370 [hep-ph]].
  C.~Hagedorn, S.~F.~King and C.~Luhn,
  JHEP {\bf 1006}, 048 (2010)
  [arXiv:1003.4249 [hep-ph]].
  R.~de Adelhart Toorop, F.~Bazzocchi and L.~Merlo,
  JHEP {\bf 1008}, 001 (2010)
  [arXiv:1003.4502 [hep-ph]];
  G.~J.~Ding,
  arXiv:1006.4800 [hep-ph];
  Y.~Daikoku and H.~Okada,
  arXiv:1008.0914 [hep-ph].







\bibitem{S3}
  P.~F.~Harrison and W.~G.~Scott,
  Phys.\ Lett.\  B {\bf 557}, 76 (2003)
  [arXiv:hep-ph/0302025];
  J.~Kubo, A.~Mondragon, M.~Mondragon and E.~Rodriguez-Jauregui,
  Prog.\ Theor.\ Phys.\  {\bf 109}, 795 (2003)
  [Erratum-ibid.\  {\bf 114}, 287 (2005)]
  [arXiv:hep-ph/0302196].
  T.~Kobayashi, J.~Kubo and H.~Terao,
  Phys.\ Lett.\  B {\bf 568}, 83 (2003)
  [arXiv:hep-ph/0303084];
  S.~L.~Chen, M.~Frigerio and E.~Ma,
  Phys.\ Rev.\  D {\bf 70}, 073008 (2004)
  [Erratum-ibid.\  D {\bf 70}, 079905 (2004)]
  [arXiv:hep-ph/0404084];
  F.~Caravaglios and S.~Morisi,
  arXiv:hep-ph/0503234;
  S.~Morisi and M.~Picariello,
  Int.\ J.\ Theor.\ Phys.\  {\bf 45}, 1267 (2006)
  [arXiv:hep-ph/0505113];
  Y.~Koide,
  Phys.\ Rev.\  D {\bf 73}, 057901 (2006)
  [arXiv:hep-ph/0509214].
  N.~Haba and K.~Yoshioka,
  Nucl.\ Phys.\  B {\bf 739}, 254 (2006)
  [arXiv:hep-ph/0511108];
  N.~Haba, A.~Watanabe and K.~Yoshioka,
  Phys.\ Rev.\ Lett.\  {\bf 97}, 041601 (2006)
  [arXiv:hep-ph/0603116];
  R.~N.~Mohapatra, S.~Nasri and H.~B.~Yu,
  Phys.\ Lett.\  B {\bf 639}, 318 (2006)
  [arXiv:hep-ph/0605020];
  Y.~Koide,
  Eur.\ Phys.\ J.\  C {\bf 50}, 809 (2007)
  [arXiv:hep-ph/0612058];
 A.~Mondragon, M.~Mondragon and E.~Peinado,
  Phys.\ Rev.\  D {\bf 76}, 076003 (2007)
  [arXiv:0706.0354 [hep-ph]];
  C.~Y.~Chen and L.~Wolfenstein,
  Phys.\ Rev.\  D {\bf 77}, 093009 (2008)
  [arXiv:0709.3767 [hep-ph]];
  M.~Mitra and S.~Choubey,
  Phys.\ Rev.\  D {\bf 78}, 115014 (2008)
[arXiv:0806.3254 [hep-ph]];
  D.~A.~Dicus, S.~F.~Ge and W.~W.~Repko,
  Phys.\ Rev.\  D {\bf 82}, 033005 (2010)
  [arXiv:1004.3266 [hep-ph]].


\bibitem{S3xA4}
  K.~S.~Babu and S.~Gabriel,
  arXiv:1006.0203 [hep-ph].


\bibitem{PSL27}
  S.~F.~King and C.~Luhn,
  Nucl.\ Phys.\  B {\bf 820}, 269 (2009)
  [arXiv:0905.1686 [hep-ph]].
  S.~F.~King and C.~Luhn,
  Nucl.\ Phys.\  B {\bf 832}, 414 (2010)
  [arXiv:0912.1344 [hep-ph]];
  R.~Zwicky and T.~Fischbacher,
  Phys.\ Rev.\  D {\bf 80}, 076009 (2009)
  [arXiv:0908.4182 [hep-ph]].





\bibitem{Quat}
  P.~H.~Frampton and O.~C.~W.~Kong,
  Phys.\ Rev.\  D {\bf 53}, 2293 (1996)
  [arXiv:hep-ph/9511343];
  P.~H.~Frampton and A.~Rasin,
  Phys.\ Lett.\  B {\bf 478}, 424 (2000)
  [arXiv:hep-ph/9910522]; 
  M.~Frigerio, S.~Kaneko, E.~Ma and M.~Tanimoto,
  Phys.\ Rev.\  D {\bf 71}, 011901 (2005)
  [arXiv:hep-ph/0409187];
  M.~Frigerio,
  arXiv:hep-ph/0505144;
  K.~S.~Babu and J.~Kubo,
  Phys.\ Rev.\  D {\bf 71}, 056006 (2005)
  [arXiv:hep-ph/0411226];
  M.~Frigerio and E.~Ma,
  Phys.\ Rev.\  D {\bf 76}, 096007 (2007)
  [arXiv:0708.0166 [hep-ph]].



  
\bibitem{dihedral}
  C.~D.~Carone and R.~F.~Lebed,
  Phys.\ Rev.\  D {\bf 60}, 096002 (1999)
  [arXiv:hep-ph/9905275];
  W.~Grimus and L.~Lavoura,
  Phys.\ Lett.\  B {\bf 572}, 189 (2003)
  [arXiv:hep-ph/0305046];
W.~Grimus, A.~S.~Joshipura, S.~Kaneko, L.~Lavoura and M.~Tanimoto,
  JHEP {\bf 0407}, 078 (2004)
  [arXiv:hep-ph/0407112];
  E.~Ma,
  Fizika B {\bf 14}, 35 (2005)
  [arXiv:hep-ph/0409288];
  S.~L.~Chen and E.~Ma,
  Phys.\ Lett.\  B {\bf 620}, 151 (2005)
  [arXiv:hep-ph/0505064];
  C.~Hagedorn, M.~Lindner and F.~Plentinger,
  Phys.\ Rev.\  D {\bf 74}, 025007 (2006)
  [arXiv:hep-ph/0604265];
  Y.~Kajiyama, J.~Kubo and H.~Okada,
  Phys.\ Rev.\  D {\bf 75}, 033001 (2007)
  [arXiv:hep-ph/0610072];
  F.~Feruglio, C.~Hagedorn, Y.~Lin and L.~Merlo,
  Nucl.\ Phys.\  B {\bf 775}, 120 (2007)
  [Erratum-ibid.\  {\bf 836}, 127 (2010)]
  [arXiv:hep-ph/0702194];
  P.~Ko, T.~Kobayashi, J.~h.~Park and S.~Raby,
  Phys.\ Rev.\  D {\bf 76}, 035005 (2007)
  [Erratum-ibid.\  D {\bf 76}, 059901 (2007)]
  [arXiv:0704.2807 [hep-ph]];
  A.~Blum, C.~Hagedorn and M.~Lindner,
  Phys.\ Rev.\  D {\bf 77}, 076004 (2008)
  [arXiv:0709.3450 [hep-ph]];
H.~Ishimori, T.~Kobayashi, H.~Ohki, Y.~Omura, R.~Takahashi and M.~Tanimoto,
Phys. Lett. B 662 (2008) 178 [arXiv:0802.2310 [hep-ph]];
H.~Ishimori, T.~Kobayashi, H.~Ohki, Y.~Omura, R.~Takahashi and M.~Tanimoto,
Phys. Rev. D 77 (2008) 115005 [arXiv:0803.0796 [hep-ph]];
  A.~Adulpravitchai, A.~Blum and C.~Hagedorn,
  JHEP {\bf 0903}, 046 (2009)
  [arXiv:0812.3799 [hep-ph]];
  C.~Hagedorn and R.~Ziegler,
  arXiv:1007.1888 [hep-ph].

\bibitem{A5}
  L.~L.~Everett and A.~J.~Stuart,
  Phys.\ Rev.\  D {\bf 79}, 085005 (2009)
  [arXiv:0812.1057 [hep-ph]].


\bibitem{D10}
  A.~Adulpravitchai, A.~Blum and W.~Rodejohann,
  New J.\ Phys.\  {\bf 11}, 063026 (2009)
  [arXiv:0903.0531 [hep-ph]].

\bibitem{datta}
  A.~Datta, F.~S.~Ling and P.~Ramond,
  Nucl.\ Phys.\  B {\bf 671}, 383 (2003)
  [arXiv:hep-ph/0306002].



  
  \bibitem{Chen:2010ty}
  C.~S.~Chen, T.~W.~Kephart and T.~C.~Yuan,
  arXiv:1011.3199 [hep-ph].
  
  \bibitem{reviews}
  G.~Altarelli and F.~Feruglio,
  New J.\ Phys.\  {\bf 6}, 106 (2004)
  [arXiv:hep-ph/0405048];
  R.~N.~Mohapatra {\it et al.},
  Rept.\ Prog.\ Phys.\  {\bf 70}, 1757 (2007)
  [arXiv:hep-ph/0510213]; 
  A.~Strumia and F.~Vissani,
  arXiv:hep-ph/0606054.
  E.~Ma,
  arXiv:0705.0327 [hep-ph].
  G.~Altarelli,
  arXiv:0705.0860 [hep-ph].
  G.~Altarelli,
  arXiv:0711.0161 [hep-ph].
  F.~Feruglio, C.~Hagedorn, Y.~Lin and L.~Merlo,
  arXiv:0808.0812 [hep-ph];
  G.~Altarelli,
  arXiv:0905.2350 [hep-ph].
  G.~Altarelli,
  Nuovo Cim.\  C {\bf 32N5-6}, 91 (2009)
  [arXiv:0905.3265 [hep-ph]];
  E.~Ma,
  arXiv:0908.1770 [hep-ph];
  S.~F.~King,
  AIP Conf.\ Proc.\  {\bf 1200}, 103 (2010)
  [arXiv:0909.2969 [hep-ph]];
  G.~Altarelli and F.~Feruglio,
  arXiv:1002.0211 [hep-ph];
  H.~Ishimori, T.~Kobayashi, H.~Ohki, H.~Okada, Y.~Shimizu and M.~Tanimoto,
  Prog.\ Theor.\ Phys.\ Suppl.\  {\bf 183}, 1 (2010)
  [arXiv:1003.3552 [hep-th]];
  M.~Zralek,
  Acta Phys.\ Polon.\  B {\bf 41}, 1477 (2010).
  
  \bibitem{rodejohann}
  C.~H.~Albright, A.~Dueck and W.~Rodejohann,
  arXiv:1004.2798 [hep-ph].
  
 \bibitem{ludl}
 P.~O.~Ludl,
  arXiv:0907.5587 [hep-ph];
  W.~Grimus and P.~O.~Ludl,
  J.\ Phys.\ A  {\bf 43}, 445209 (2010)
  [arXiv:1006.0098 [hep-ph]].;
  P.~O.~Ludl,
  J.\ Phys.\ A  {\bf 43}, 395204 (2010)
  [arXiv:1006.1479 [math-ph]].
  


\bibitem{discretequarks}
 P.~H.~Frampton and O.~C.~W.~Kong,
  Phys.\ Rev.\ Lett.\  {\bf 75}, 781 (1995)
  [arXiv:hep-ph/9502395];
  A.~Blum, C.~Hagedorn and A.~Hohenegger,
  JHEP {\bf 0803}, 070 (2008)
  [arXiv:0710.5061 [hep-ph]]; 
  A.~Blum and C.~Hagedorn,
  Nucl.\ Phys.\  B {\bf 821}, 327 (2009)
  [arXiv:0902.4885 [hep-ph]].


\bibitem{u2}
  R.~Barbieri, G.~R.~Dvali and L.~J.~Hall,
  Phys.\ Lett.\  B {\bf 377}, 76 (1996)
  [arXiv:hep-ph/9512388];
  R.~Barbieri, L.~J.~Hall, S.~Raby and A.~Romanino,
  Nucl.\ Phys.\  B {\bf 493}, 3 (1997)
  [arXiv:hep-ph/9610449];
  A.~Romanino,
  arXiv:hep-ph/9701349.
  R.~Barbieri, L.~J.~Hall and A.~Romanino,
  Phys.\ Lett.\  B {\bf 401}, 47 (1997)
  [arXiv:hep-ph/9702315];
  C.~D.~Carone and L.~J.~Hall,
  Phys.\ Rev.\  D {\bf 56}, 4198 (1997)
  [arXiv:hep-ph/9702430];
  M.~Tanimoto,
  Phys.\ Rev.\  D {\bf 57}, 1983 (1998)
  [arXiv:hep-ph/9706497];
  G.~Eyal,
  Phys.\ Lett.\  B {\bf 441}, 191 (1998)
  [arXiv:hep-ph/9807308];
  L.~J.~Hall and N.~Weiner,
  Phys.\ Rev.\  D {\bf 60}, 033005 (1999)
  [arXiv:hep-ph/9811299];
  R.~Barbieri, L.~Giusti, L.~J.~Hall and A.~Romanino,
  Nucl.\ Phys.\  B {\bf 550}, 32 (1999)
  [arXiv:hep-ph/9812239];
  R.~Barbieri, P.~Creminelli and A.~Romanino,
  Nucl.\ Phys.\  B {\bf 559}, 17 (1999)
  [arXiv:hep-ph/9903460];
  T.~Blazek, S.~Raby and K.~Tobe,
  Phys.\ Rev.\  D {\bf 62}, 055001 (2000)
  [arXiv:hep-ph/9912482];
  A.~Aranda, C.~D.~Carone and P.~Meade,
  Phys.\ Rev.\  D {\bf 65}, 013011 (2002)
  [arXiv:hep-ph/0109120];
  S.~Gopalakrishna and C.~P.~Yuan,
  arXiv:hep-ph/0402096.
  Y.~Kobashi and A.~Sugamoto,
  arXiv:hep-th/0407118.
  S.~Gopalakrishna and C.~P.~Yuan,
  Phys.\ Rev.\  D {\bf 71}, 035012 (2005)
  [arXiv:hep-ph/0410181];
  Z.~Han,
  Phys.\ Rev.\  D {\bf 73}, 015005 (2006)
  [arXiv:hep-ph/0510125];
  T.~Goto, Y.~Okada, T.~Shindou and M.~Tanaka,
  Phys.\ Rev.\  D {\bf 77}, 095010 (2008)
  [arXiv:0711.2935 [hep-ph]].

\bibitem{HPS}
  P.~F.~Harrison, D.~H.~Perkins and W.~G.~Scott,
  Phys.\ Lett.\  B {\bf 530}, 167 (2002)
  [arXiv:hep-ph/0202074].

  
  \bibitem{alexlisainprep}  
  L.~L.~Everett and A.~J.~Stuart, in preparation.
  
  \bibitem{backhousegard}
N.~B.~Backhouse and P.~Gard, J.\ Phys. \ A {\bf 7} 2101 (1974).  
  
\bibitem{cumminspatera}
C.~J.~Cummins and J.~Patera, J.\ Math. \ Phys. {\bf 29}, 1736 (1988).

\bibitem{shirai}
  K.~Shirai,
J. \ Phys. \ Soc.\ Jpn.\ {\bf 61} 2735 (1992).
  
\bibitem{hoyle}
 R.~B.~Hoyle, 
 Physica D {\bf 191}, 261(2004).  

\bibitem{luhn}
  C.~Luhn, S.~Nasri and P.~Ramond,
  J.\ Math.\ Phys.\  {\bf 48}, 123519 (2007)
  [arXiv:0709.1447 [hep-th]];
  C.~Luhn and P.~Ramond,
  J.\ Math.\ Phys.\  {\bf 49}, 053525 (2008)
  [arXiv:0803.0526 [hep-th]].



\bibitem{Coxeter}
  H.~Coxeter and W.~Moser,
  {\it Generators and Relations for Discrete Groups},  Springer-Verlag, Berlin (1957).
  
\bibitem{Lomont}
  J.~S.~Lomont,  {\it Applications of Finite Groups}, Academic Press, New York (1959).
  
 \bibitem{Hamermesh}
  M.~Hamermesh, {\it Group Theory and its Applications to Physical Problems}, Addison-Wesley, Reading (1964).

\bibitem{Ramond}
P.~Ramond, {\it Group Theory: A Physicist's Survey}, Cambridge University Press (2010).  
  



\end{thebibliography}
\end{document}